\documentclass[RNAAS]{aastex63}

\begin{document}

\title{GN-z11-flash in the context of Gamma-Ray Burst Afterglows}

\correspondingauthor{D. A. Kann}
\email{kann@iaa.es}

\author[0000-0003-2902-3583]{D. A. Kann}
\affiliation{IAA-CSIC, Glorieta de la Astronom\'ia, s/n, 18008 Granada, Spain}

\author[0000-0003-0034-7891]{M. Blazek}
\affiliation{IAA-CSIC, Glorieta de la Astronom\'ia, s/n, 18008 Granada, Spain}

\author[0000-0001-7717-5085]{A. de Ugarte Postigo}
\affiliation{IAA-CSIC, Glorieta de la Astronom\'ia, s/n, 18008 Granada, Spain}
\affiliation{DARK, Niels Bohr Institute, University of Copenhagen, Lyngbyvej 2, DK-2100 Copenhagen
{\O}, Denmark}

\author[0000-0002-7978-7648]{C. C. Th\"one}
\affiliation{IAA-CSIC, Glorieta de la Astronom\'ia, s/n, 18008 Granada, Spain}


\begin{abstract}
The recently discovered rapid transient GN-z11-flash has been suggested to be the prompt-emission ultraviolet flash associated with a gamma-ray burst serendipitously exploding in the ultra-high-$z$ galaxy GN-z11. We here place the flash into the context of the early ultraviolet emission of gamma-ray bursts, and find it is in agreement with the luminosity distribution of these events.
\end{abstract}

\keywords{High-redshift galaxies (734), Gamma-ray bursts (629)}

\section{}

Gamma-Ray Bursts (GRBs) are among the most powerful explosions in the Universe, and their optical emission at very early times can be dominated by extremely luminous ultraviolet (UV) flashes \citep{Akerlof1999Nature,Racusin2008Nature,Bloom2009ApJ,Vestrand2014Science}, which may be directly linked to the higher-energy prompt emission \citep[e.g.,][]{Blake2005Nature,Vestrand2005Nature}. Such prompt flashes are the most luminous UV/optical sources known \citep{Kann2007AJ,Perley2011AJ}, implying GRBs are detectable across a broad wavelength range up to the highest studied redshifts \citep{Tanvir2009Nature,Salvaterra2009Nature,Cucchiara2011ApJ,Tanvir2018ApJ}. This makes them highly interesting probes of cosmology, and the main target of mission concepts such as the Gamov Explorer \citep{White2020arXiv} and THESEUS \citep{Amati2018AdSpR}.

Recently, \cite{Jiang2020NA1} reported a spectroscopic redshift for the ultra-high-$z$ candidate galaxy GN-z11 \citep{Oesch2016ApJ}, finding it to lie at $z=10.957$, confirming it to be the most distant object known at the time. Serendipitously, as \citet[][J20 hereafter]{Jiang2020NA2} report, a single (2D spectrum) image of their 5.3 hr $K$-band spectrum contains a strong signal that is not seen in the previous or the subsequent integration. They exclude atmospheric, terrestrial, and Solar System sources and present evidence that the transient is very likely associated with GN-z11 and therefore at the same redshift. This implies extreme luminosity, and they therefore argue that GN-z11-flash is most likely the prompt flash of a GRB exploding while they were observing the galaxy. It is extremely unlikely that such an incidence would occur, however, all other explanations J20 present are even more unlikely, mostly significantly so.

Assuming GN-z11-flash is indeed associated with the $z\approx11$ galaxy, we here compare its luminosity with a large sample of GRB afterglows \citep[][Kann et al. 2021a,b, in prep.]{Kann2006ApJ,Kann2010ApJ,Kann2011ApJ}. J20 report the redshift ($z=10.957$), a lack of dust extinction, and the spectral slope they measure from their calibrated spectrum ($\beta=-1.2\pm0.4$ assuming $F_\nu\propto\nu^{-\beta}$). Following J20, we here assume the emission is synchrotron radiation and this spectral slope can be extrapolated. We neglect Galactic foreground extinction, as this galaxy is in the GOODS/Hubble Deep Field North, and it is negligible (a few millimag). We use a $\lambda$CDM concordance cosmology with $H_0=72$ km/s/Mpc, $\Omega_M=0.27$, and $\Omega_\lambda=0.73$. J20 report a slit-loss corrected flux density in the $K$ band of $0.1-0.2$ mJy, we assume $0.15\pm0.05$ mJy. This corresponds to $K=18.46^{+0.44}_{-0.31}$ mag (AB system, $K=16.59$ mag in Vega). From the spectral slope, we find $K-R_C=1.54$ mag (AB; $K-R_C=-0.15$ mag Vega). Using the redshift and spectral slope, we find a magnitude correction to $z=1$ of $dRc=-1.98^{+0.78}_{-0.76}$ mag (the blue spectrum and the large distance partially cancel each other out) following \cite{Kann2006ApJ}. In total, this yields $R_C=14.46^{+0.90}_{-0.82}$ mag at $z=1$, Vega system. Following J20, the non-detection in the following spectrum implies a flux density at least ten times fainter, giving us a corresponding $R_C>17.0$ mag. J20 did not find a coincident GRB detection. We assume the GRB began ($t=T_0$) with the end of the previous spectral integration, this yields logarithmic central points of 14.0 s for the detection and 53.9 s for the upper limit, in the $z=1$ frame. For these times, the decay slope is $\alpha\gtrsim1.7$ ($F(t)\propto t^{-\alpha}$), a reasonable value for early GRB flashes.

\begin{figure}[ht!]
\begin{center}
\includegraphics[scale=1,angle=0]{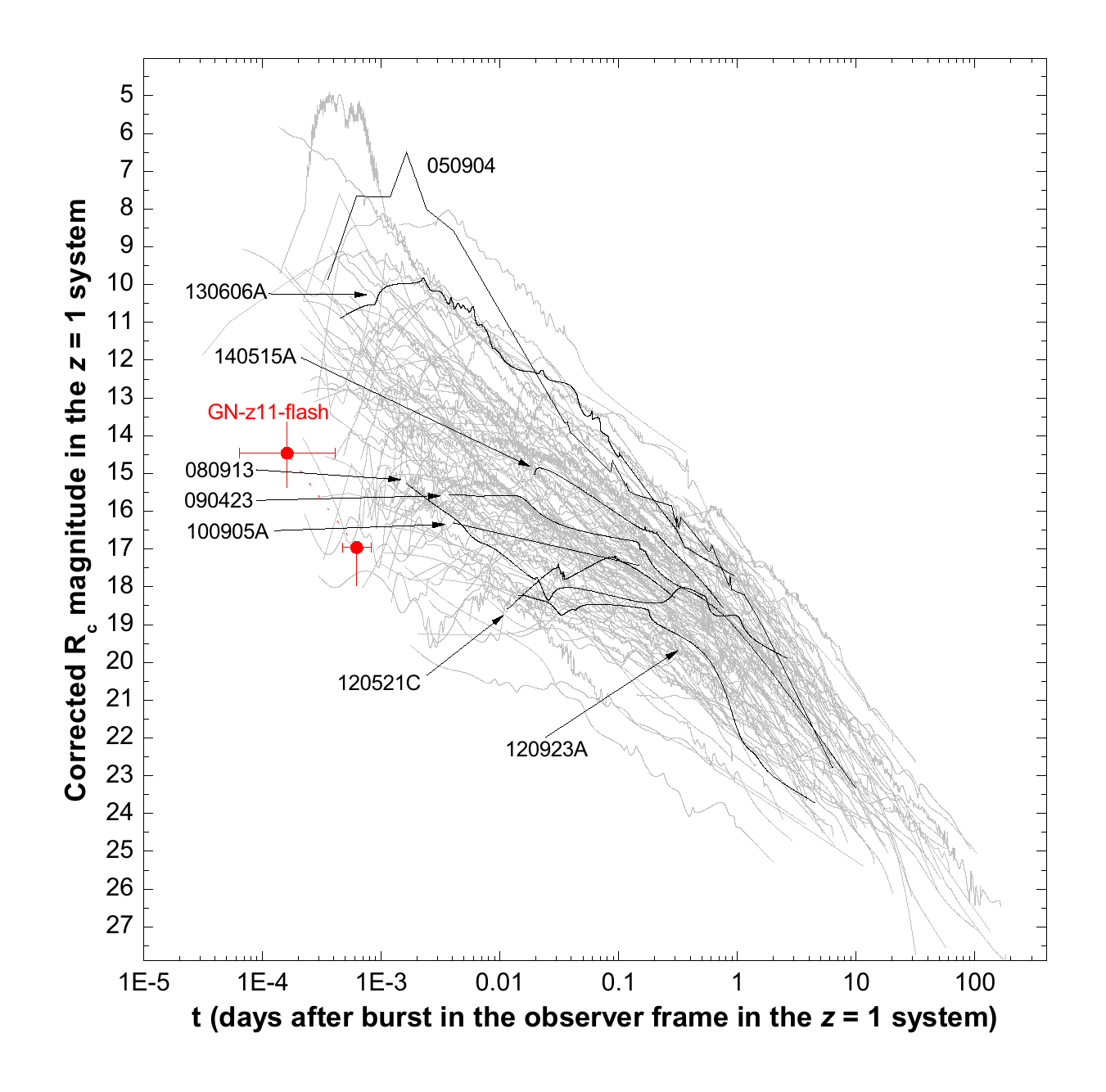}
\caption{GN-z11-flash (red) compared to a large sample of GRB afterglows shifted to $z=1$. We also highlight afterglows of high-redshift GRBs at $z>6$. GN-z11-flash is seen to agree with the early luminosity distribution of GRB afterglows.}
\end{center}
\label{fig}
\end{figure}

We plot the final result (red) in Fig. \ref{fig}, highlighting a sample of $z>6$ GRB afterglows (Kann et al. 2021a, in prep.). It can clearly be seen that GN-z11-flash is in full accordance with the luminosity distribution of early GRB optical emission, supporting the interpretation of J20. Considering no further emission was detected in subsequent spectra, we even infer this was a quite faint afterglow. In the sample of Kann et al. (2021a, in prep.), we find 70 GRB afterglows with detections at the time of the upper limit from J20 (assuming our $T_0$; note some are extrapolations), 63/70 (90\%) are brighter than the upper limit of $R_C>17$ mag.

\acknowledgments

DAK acknowledges support from the Spanish National Research Project RTI2018-098104-J-I00 (GRBPhot). MB acknowledges funding associated to a personal t\'ecnico de apoyo fellowship (PTA2016-13192-I). AdUP and CCT acknowledge support from Ram\'on y Cajal fellowships RyC-2012-09975 and RyC-2012-09984. 







\end{document}